\documentclass[aps,pre,
,floatfix,footinbib,preprint]{revtex4-1}
\usepackage{epsfig}
\usepackage{graphicx}% Include figure files
\usepackage{dcolumn}% Align table columns on decimal point
\usepackage{bm}% bold math
\usepackage{mathrsfs}
\usepackage{amsmath}
\usepackage{physics}
\usepackage{bbold}
\usepackage{color}
\usepackage{epstopdf}
\usepackage[colorlinks=true,pdfborder=001,linkcolor=blue,anchorcolor=blue,citecolor=blue,urlcolor=blue]{hyperref}

\usepackage{appendix}

\definecolor{indiagreen}{rgb}{0.07, 0.53, 0.03}
\definecolor{indianyellow}{rgb}{0.89, 0.66, 0.34}
\begin{document}

\title{Non-thermal vibrations in biased molecular junctions}%

\author{Tao Wang}
\author{Lei-Lei Nian}
\email{llnian@hust.edu.cn}
\author{Jing-Tao L\"u}
\email{jtlu@hust.edu.cn}
\affiliation{School of Physics and Wuhan National High Magnetic Field Center, Huazhong University of Science and Technology, Wuhan
430074, P. R. China}

\date{\today}% It is always \today, today,
\begin{abstract}
We study vibrational statistics in current-carrying model molecular junctions using master equation approach. Especially, we concentrate on the validity of using an effective temperature $T_{\rm eff}$ to characterize the nonequilibrium steady state of a vibrational mode. We identify cases where a single $T_{\rm eff}$ can not fully describe one vibrational state. In such cases, the probability distribution among different vibrational states does not follow the Boltzmann type. Consequently, the actual entropy (free energy) of the vibrational mode is lower (higher) than the corresponding thermal value given by $T_{\rm eff}$, indicating extra work can be extracted from these states. Our results will be useful for the study of non-thermal vibrational state in thermodynamics of nanoscale systems, and its usage in nanoscale heat engines.

%\begin{description}
%05.70.Ln, 05.20.-y, 05.70.-a
%\end{description}
\end{abstract}

%\pacs{73.63.kv, 73.23.-b, 71.38.-k}% PACS, the Physics and Astronomy
\maketitle

\section{Introduction}
In recent years, electron transport through single molecular junction has received considerable attention both experimentally and theoretically in view of its importance in molecular electronics\cite{
%heath2003molecular,
nitzan2003electron,flood2004whence,xiang2016molecular,xin2019concepts,gehring2019single,thoss2018}. Many techniques have been developed to couple a single molecule to two electrodes, and to measure its electrical conductance\cite{reed1997conductance,stipe1998single,yu2004inelastic,elbing2005single}. The conductance is not only affected by the molecule in the junction, but also, by the coupling between the molecule and the electrodes, the electric structure of the electrodes, and the interaction between electrons and molecular vibrations\cite{galperin2007molecular}. The vibrations can be excited when the applied voltage bias exceeds the molecular vibrational energy. Thus, energy transfer from the electronic to the vibrational degrees of freedom takes place, resulting in energy accumulation in the vibrational system and resultant heat transport\cite{pecchia07,ness2005,wang2008quantum,lu2015effects,dubi2011colloquium,li2012phononics}. This is loosely termed Joule heating, although deterministic energy transfer through work may take place simultaneously\cite{dundas2009,lu2010blowing,bode2011}. This may in turn lead to the conformation change and atomic rearrangements\cite{park2000nanomechanical,gaudioso2000vibrationally}.  In the extreme case, the molecular junction can be destroyed through breaking of chemical bond. On the other hand, through specially designed electronic structure, one may use the non-equilibrium effect to cool the molecular junctions, leading to current-induced cooling\cite{galperin2009cooling,hartle2011resonant,simine2012vibrational,romano2010heating,hartle2018cooling}. 
 
Theoretically, the concept of effective temperature has been used to describe the junction heating and cooling when it reaches the nonequilibrium steady state under applied voltage bias\cite{galperin2007heat,galperin2007molecular,huang2007local,zhang2019local}. It describes the statistical properties of a vibrational mode. The purpose of this work is to show that this is not always the case. We illustrate non-thermal statistical properties of the vibrations by considering two model systems that have been widely used in previous studies. In the first model, we consider a vibrational laser where one vibrational mode couples to two electronic states via the Su-Schrieffer-Heeger-like coupling\cite{lu2011laserlike,simine2012vibrational,foti2018,nitzan2018kinetic}.
In the second model, we consider Holstein-type on-site coupling between one electronic state with one vibrational mode\cite{braig2003,mitra2004,galperin2006resonant}. In both models, we find situations where one effective temperature is not enough to describe the statistical properties of the vibrational mode.

\section{Models and methods}
\subsection{Model I: A two-level molecular junction}
\label{Model-I}

\begin{figure}
\includegraphics[scale=0.4]{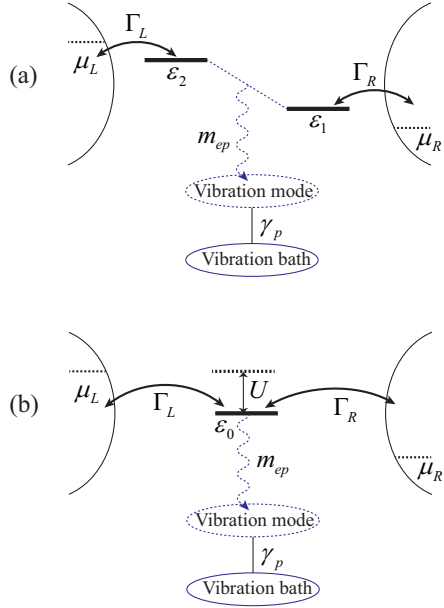}
\caption{(Color online)  (a) Schematic model of the transport in a bias-driven single molecular junction. The molecule consisting of two levels $\varepsilon_{1}$ and $\varepsilon_{2}$ is coupled to two electrodes (L and R) characterized by energy independent parameters $\Gamma_{L}$ and  $\Gamma_{R}$. The vibrational mode can be excited due to electron-vibration $m_{ep}$ when the bias voltage ($eV_{bais}=\mu_{L}-\mu_{R}$) between electrodes is large than the energy of the mode. The statistics of the vibrational mode can be obtained from the bath with a dissipation ratio $\gamma_{p}$. (b) Schematic representation of a single-level molecular junction similar to (a). Here, the vibrational excitation is caused by the Holstein-type on-site coupling between one electronic state $\varepsilon_{0}$ with Coulomb interaction $U$.}
\label{model}
\end{figure}

The first model we consider is a molecular junction consisting of two levels coupled to electrodes as depicted in Fig.~\ref{model}(a). The vibrational mode can be excited by the inelastic transitions between two electronic states. 
The corresponding Hamiltonian is
\begin{equation}
\begin{split}
&\mathcal{H}=\mathcal{H}_{m}+\mathcal{H}_{el}+\mathcal{H}_{ep}+\mathcal{H}_{p}+\mathcal{H}_{b},\\
&\mathcal{H}_{m}= \sum_{i=1,2} \varepsilon_{i} n_i + U_{12}n_{1}n_{2},\\
&\mathcal{H}_{el}=\sum_{\alpha=L,R}\sum_{k}(\varepsilon_{k\alpha}-\mu_{\alpha})c_{k\alpha}^{\dagger}c_{k\alpha}+\sum_{\alpha=L,R}\sum_{k}\sum_{i=1,2}(V_{\alpha k,i}c_{k\alpha}^{\dagger}d_i + h.c.),\\
&\mathcal{H}_{ep}=m_{ep}(a_{p}^{\dagger}d_1^\dagger d_2+a_{p}d_2^\dagger d_1),\\
&\mathcal{H}_{p}=\hbar\omega_{p}(a_{p}^{\dagger}a_{p}+\frac{1}{2}),\\
&\mathcal{H}_{b}=\sum_{\alpha}\hbar\omega_{\alpha}(a_{\alpha}^{\dagger}a_{\alpha}+\frac{1}{2})+\sum_{\alpha}t_{\alpha p}(a_{\alpha}^{\dagger}+a_{\alpha})(a_{p}^{\dagger}+a_{p}),
\end{split}
\label{model-I-Hamiltnoian}
\end{equation}
where $\mathcal{H}_{m}$ is the Hamiltonian of the molecule, $ n_i = d_i^\dagger d_i$ is the electron number operator for state $i$, $\varepsilon_i$ is the corresponding energy, and $U_{12}$ is the inter-site Coulomb charging energy. The two electrodes and their coupling with molecule are described by $\mathcal{H}_{el}$. 
$c_{k\alpha}^{\dagger}~(c_{k\alpha})$ is the creation (annihilation) operator of an electron with the wave vector $k$ in the electrode $\alpha$. $\varepsilon_{k\alpha}$ and $\mu_{\alpha}$ are the corresponding energy and the chemical potential, respectively. $V_{\alpha k}$ is the electrode-molecule coupling parameter.
The electronic states couple to a vibrational mode, $\mathcal{H}_{ep}$ is the corresponding Hamiltonian and the vibrational mode is described by $\mathcal{H}_{p}$. The last term $\mathcal{H}_{b}$ describes damping of the vibrational mode due to coupling to a vibrational bath. 
$a_{p}^{\dagger}~(a_{p})$ and $a_{\alpha}^{\dagger}~(a_{\alpha})$ are the creation (annihilation) operators of the vibrational mode and the bath with angular frequencies $\omega_{p}$ and $\omega_{\alpha}$, with $t_{\alpha p}$
 being the their coupling.

To study the vibration statistics, we use the master equation approach of the Lindblad form. The molecule-electrodes coupling are regarded as a  perturbation\cite{breuer2002theory,scully1997quantum}. 
We furthermore consider the molecule system in the strong Coulomb blockade
regime ($U_{12} \rightarrow \infty$), that is, only the occupation by a single excess
electron is allowed. Then, the effective Hilbert space of the molecular system
is spanned by three states, which are $|0 \rangle=|0,0 \rangle$, $|a \rangle=|1,0 \rangle$, and $|b \rangle=|0,1 \rangle$.
Meanwhile, we can define creation operators of the ground and excited states for the molecule as
$d_{g}^{\dagger}=|a \rangle  \langle0|$ and $d_{e}^{\dagger}=|b\rangle \langle0| $ with energies $\varepsilon_{1}$ and $\varepsilon_{2}$, respectively. The Hamiltonian in Eq.~\ref{model-I-Hamiltnoian} can be rewritten in such representation.
Under the Born-Markov approximation,  the reduced density matrix for electron-vibration system follows the following equation of motion
\begin{equation}
\dot{\rho}=\frac{1}{i\hbar}[\mathcal{H}_{0},\rho]+\mathcal{L}_{el}[\rho]+\mathcal{L}_{p}[\rho],
\label{Lindblad}
\end{equation}
with  $\mathcal{H}_{0}=\mathcal{H}_{m}+\mathcal{H}_{p}+\mathcal{H}_{ep}$. The first term at the right hand side describes the quantum coherent evolution of electron-vibration system. 
The last two terms correspond to the dissipation of the system due to the interaction with electrodes and vibrational bath. We have
\begin{equation}
\begin{split}
\mathcal{L}_{el}[\rho]&=\frac{1}{2}\sum_{\alpha }\Gamma_{\alpha 1}\bigg\{f_{\alpha}(\varepsilon_{g})\mathcal{D}[d_{g},\rho]+(1-f_{\alpha}(\varepsilon_{g}))\mathcal{D}[d_{g}^{\dagger},\rho]\bigg\}\\
&+\frac{1}{2}\sum_{\alpha}\Gamma_{\alpha 2}\bigg\{f_{\alpha}(\varepsilon_{e})\mathcal{D}[d_{e},\rho]+(1-f_{\alpha}(\varepsilon_{e}))\mathcal{D}[d_{e}^{\dagger},\rho]\bigg\},
\end{split}
\end{equation}
where $\Gamma_{\alpha i}(\varepsilon)=2\pi\sum_{k}V_{\alpha k,i}^{2} \delta(\varepsilon-\varepsilon_{k\alpha})$ is the level broadening function of the state $i$ due to coupling with electrode $\alpha$. We have ignored its energy dependence here. The Lindblad superoperators act according to $\mathcal{D}[\mathcal{A},\rho]=2\mathcal{A}^{\dagger}\rho \mathcal{A}-\{\mathcal{A}
\mathcal{A}^{\dagger},\rho\}$. For the vibration $\mathcal{L}_{p}[\rho]$ can be written as
\begin{equation}
\mathcal{L}_{p}[\rho]=\frac{\gamma_{p}}{2}(1+n_{B})\mathcal{D}[a_{p}^{\dagger},\rho]+\frac{\gamma_{p}}{2}n_{B}\mathcal{D}[a_{p},\rho],
\end{equation}
where $n_{B}=[e^{\hbar\omega_{p}/k_{B}T}-1]^{-1}$ is the average occupation of the vibrational mode $\omega_{p}$ in equilibrium state at temperature $T$.

Using the standard quantum master equation procedure, the time evolution of the vibrational density matrix element can be written as
\begin{equation}
\begin{split}
\frac{dp_{m,n}}{dt}&=-i\omega_{p}(m-n)p_{m,n}-im_{ep}\bigg[\sqrt{m+1}\rho_{m+1,n}^{ge}-\sqrt{n+1}\rho_{m,n+1}^{eg}+
\sqrt{m}\rho_{m-1,n}^{eg}-\sqrt{n}\rho_{m,n-1}^{ge}\bigg]\\
&+\frac{\gamma_{p}}{2}\bigg[2(n_{B}+1)\sqrt{(m+1)(n+1)}p_{m+1,n+1}-(n_{B}+1)(m+n)p_{m,n}\\
&+2n_{B}\sqrt{mn}p_{m-1,n-1}-n_{B}(m+n+2)p_{m,n}\bigg],
\end{split}
\label{pmn}
\end{equation}
where the combined density matrix elements $\rho_{mn}^{ge}$ and $\rho_{mn}^{eg}$ are given in Appendix \ref{appd:MEdo}.
For $m=n$, $p_{m,m}$ describes the probability of finding $m$ vibrational quanta.

\subsection{Model II: A single-level molecular junction}
\label{Model-II}
A single energy-level spin non-degenerate model in Fig.~\ref{model}(b) is considered in this case. The corresponding Hamiltonian is
\begin{equation}
\begin{split}
&\mathcal{H}=\mathcal{H}_{m}+\mathcal{H}_{el}+\mathcal{H}_{ep}+\mathcal{H}_{p}+\mathcal{H}_{b},\\
&\mathcal{H}_{m}=\varepsilon_{0}n + Un(n -1),\\
&\mathcal{H}_{el}= \sum_{\alpha=L,R}\sum_{k}(\varepsilon_{k\alpha}-\mu_{\alpha})c_{k\alpha}^{\dagger}c_{k\alpha} + 
\sum_{\alpha=L,R}\sum_{k}(V_{\alpha k}c_{k\alpha}^{\dagger}d + h.c.),\\
&\mathcal{H}_{ep}=m_{ep}(a_{p}^{\dagger}+a_{p})n,\\
\end{split}
\label{model-II-Hamiltnoian}
\end{equation}
where $ n = d^{\dagger}d $ is the electron occupation number operator on the molecule, $V_{\alpha k}$ is the electrode-molecule coupling parameter. The Hamiltonian for $H_{p}$ and $H_{b}$ are the same as in model I.

To consider this model, a Lang-Firsov transformation to the polaron representation can be preformed.\cite{lang1963kinetic} Applying the unitary operator $ D = e^{[\lambda (a_{p}^\dagger - a_{p}) n]} $ to the total Hamiltonian, we get
\begin{equation}
\label{eq:polaron}
\begin{split}
&\mathcal{H}' = D \mathcal{H} D^\dagger, \\
&\mathcal{H}_{m}'=(\varepsilon_{0}-m_{ep}^2\hbar\omega_{p})n + (U-2m_{ep}^2\hbar\omega_{p})n(n-1), \\
&\mathcal{H}_{el}' = \sum_{\alpha=L,R}\sum_{k}(\varepsilon_{k \alpha}-\mu_\alpha)c_{k \alpha}^{\dagger}c_{k \alpha} + \sum_{\alpha=L,R}\sum_{k}(V_{\alpha k} e^{-\lambda m_{ep}(a_{p}^\dagger - a_{p})} c_{k \alpha}^{\dagger}d + h.c.), \\
&\mathcal{H}_{p}'=\hbar\omega_{p}(a_{p}^{\dagger}a_{p}+\frac{1}{2}),\\
&\mathcal{H}_{ep}'=0, \\
&\mathcal{H}_{d}'=\mathcal{H}_{d}.
\end{split}
\end{equation}
Thus in the polaron representation, for a state $\ket{lm}$ which indicates $l$ electrons on the molecule with $m$ vibrations, we get $ \mathcal{H}' \ket{lm} = E_{lm} \ket{lm} $ with eigenvalues
\begin{equation}
E_{lm} = \varepsilon'l + U'l(l-1) + \hbar\omega_{p}(m + \frac{1}{2}),
\end{equation}
where $ \varepsilon' = \varepsilon_{0} - m_{ep}^2\hbar\omega_{p}$, $\ U' = U-2m_{ep}^2\hbar\omega_{p} $. 

In fact, a generalized master equation in this case for the reduced density operator of electron-vibration system within the Born-Markov approximation can be obtained, as shown in Eq.~\ref{Lindblad}. By using the secular approximation, we can get the evolution of vibration populations (diagonal elements) and coherences (off-diagonal elements), respectively. For our case, we mainly focus on the former, resulting in a rate equation
\begin{equation}
\begin{split}
\dot{p}_{\ket{lm}} &= \sum_{l'} \sum_{m'}\bqty{\Gamma_{(l'm')(lm)} p_{\ket{l'm'}} - \Gamma_{(lm)(l'm')} p_{\ket{lm}}}\\
    &+ m \gamma_{p} n_{\rm B} p_{\ket{l(m-1)}} + (m+1) \gamma_{p}(1+n_{\rm B}) p_{\ket{l(m+1)}}\\
    &- \bqty{(m+1) \gamma_{p} n_{\rm B} + m \gamma_{p}(1+n_{\rm B})}p_{\ket{lm}},
\end{split}
\end{equation}
where $ p_{\ket{lm}} $ is the probability that the system is in $ \ket{lm}  $ state, $ \Gamma_{(lm)(l'm')} $ is the probability that the system evolves from $ \ket{lm} $ to $ \ket{l'm'} $ and
\begin{equation}
\begin{split}
&\Gamma_{(l_{<}m)(l_{>}m')} = |M_{mm'}|^2 \sum_{\alpha = L,R} \Gamma_{\alpha} f_{\alpha}(E_{l_{>}m'} - E_{l_{<}m}) \delta_{l_{>}-l_{<},1},\\
&\Gamma_{(l_{>}m)(l_{<}m')} = |M_{mm'}|^2 \sum_{\alpha = L,R} \Gamma_{\alpha} [1-f_{\alpha}(E_{l_{>}m} - E_{l_{<}m'})] \delta_{l_{>}-l_{<},1},
\end{split}
\end{equation}
where $n_> > n_< $, and $ |M_{mm'}|^2 $ is the Franck-Condon matrix element which is presented in Appendix~\ref{appd:frME}.

By applying the steady state condition $ \dot{p}_{lm}=0 $ to the rate equations, we can calculate the probability $ p_{\ket{lm}} $. By calculating the net electron transition probability between the left electrode and the molecule, we can obtain the steady state current
\begin{equation}
I=e\sum_{lm}\sum_{l'm'} s \Gamma'_{(lm)(l'm')}p_{\ket{lm}},
\end{equation}
where the direction of the current is from the lower chemical potential side to the higher side, $e$ is the elementary charge, and $s=\pm 1$ determined by the electronic tunneling direction for a given electron transition. When an electron  tunnels from the higher chemical potential side to the lower side, $s=-1$, otherwise, $s=1$. $\Gamma'_{(lm)(l'm')}$ is a part of $\Gamma_{(lm)(l'm')}$, which gives the probability of a state transition from $\ket{lm}$ to $ \ket{l'm'} $ induced by electron tunneling between the left electrode and the molecule.

\subsection{Characteristic vibrational quantities}
We use several physical quantities to characterize the properties of vibrational state, including the average population, the effective temperature, the thermal entropy, the von Neumann entropy, and the vibration second-order coherence function.
For this we write the probability of the system with $m$ vibrational quanta as $p_{m}$, then $p_{m} = p_{m,m}$ for model I, and $p_{m} = \sum_{l} p_{\ket{lm}}$ for model II.
The average population $\langle n\rangle$ can be defined as
\begin{equation}
\begin{split}
\langle n \rangle=\sum_{m}mp_{m}.
\end{split}
\label{n}
\end{equation}
Given $\langle n\rangle$, if we assume that the vibration is in thermal equilibrium, we can define an effective temperature $T_{{\rm eff}}$ as 
\begin{equation}
\begin{split}
T_{{\rm eff}}=\frac{\hbar\omega_{p}/k_{B}}{\ln ({1}/{\langle n \rangle}+1)}.
\end{split}
\label{teff}
\end{equation}
Consequently, we can define the effective thermal entropy $S_{{\rm th}}$ as
\begin{equation}
\begin{split}
S_{{\rm th}}=k_{B}[(\langle n \rangle+1)\ln(\langle n \rangle+1)-\langle n \rangle\ln\langle n \rangle].
\end{split}
\label{Sthermal}
\end{equation}
By comparing $S_{\rm th}$ with the actual 
von Neumann entropy
\begin{equation}
\begin{split}
S_{\rm vN}=-k_{B}\sum_{m}p_{m}\ln p_{m},
\end{split}
\label{S}
\end{equation}
we can characterize the deviation from thermal state.
Another quantity we can use to quantify the non-thermal state is the vibrational second-order coherence function 
\begin{equation}
\begin{split}
g^{(2)}(0)=\frac{\langle a_{p}^{\dagger}a_{p}^{\dagger}a_{p}a_{p} \rangle}{\langle a_{p}^{\dagger}a_{p} \rangle^{2}}
=\frac{\sum_{m}m(m-1)p_{m}}{(\sum_{m}mp_{m})^{2}}.
\end{split}
\end{equation}
It has been widely used in quantum optics. One can easily verify that the vibration in thermal equilibrium yields $g^{(2)}(0)=2$. When $g^{(2)}(0)<1$, the vibration is in the anti-bunching  state, while for $g^{(2)}(0)>1$ it is in the bunching state. Thus, vibrations are bunched in thermal state due to its bosonic statistics. Moreover, when $g^{(2)}(0)=1$ the vibration is in the coherent state.

\section{Results and Discussions}
\subsection{Results for model I}
Let us begin with the case of a two-level molecular junction, where the coupling between level 2 (1) and right (left) electrode is taken as 0, that is $\Gamma_{R2}=\Gamma_{L1}=0$, see Fig.~\ref{model}(a). Such model has been used before to study resonant vibration excitation\cite{lu2011laserlike,simine2012vibrational,simine2013path,lambert2015bistable,agarwalla2015full,foti2018,nitzan2018kinetic}. The vibration in such junction is excited by the inelastic electron tunneling from level 2 to level 1. We set the Coulomb repulsion inside the molecule $U_{12}=\infty$.  The Lindblad master equation in Subsection \ref{Model-I} is used to obtain the following results.

\begin{figure}
\includegraphics[scale=0.6]{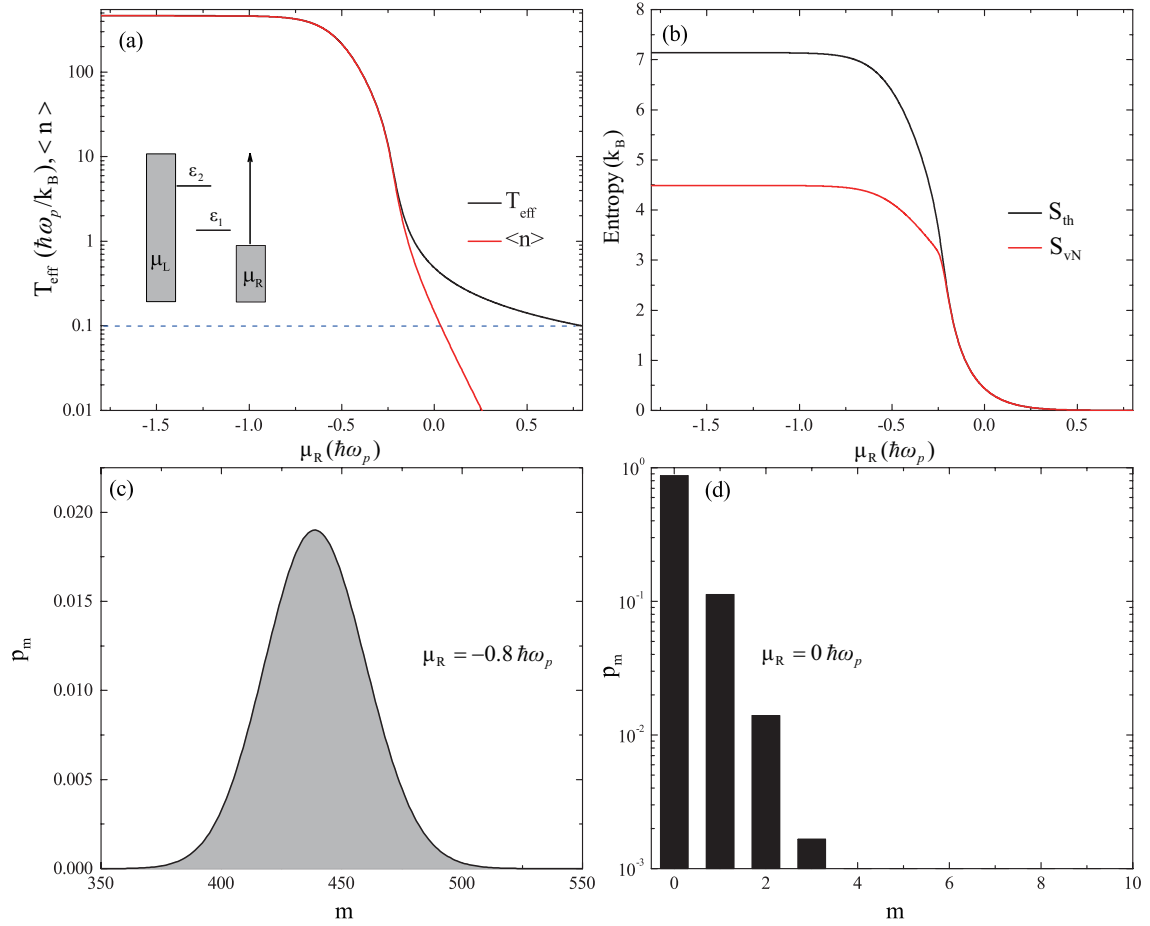}
\caption{(Color online) (a) The effective temperature $T_{\rm eff}$ and the average population $\langle n \rangle$ as a function of the chemical potential of right electrode $\mu_{R}$ with $\mu_{L}=0.8~\hbar\omega_{p}$. The inset shows two molecular levels $\varepsilon_{1}$ and $\varepsilon_{2}$ coupled to two electrodes with chemical potentials $\mu_{L}$ and $\mu_{R}$, where $\mu_{R}$ increases in the direction of the arrow. (b) The entropy vs $\mu_{R}$. (c) and (d) Vibration statistics at $\mu_{R}=0.8~\hbar\omega_{p}$ and $\mu_{R}=0~\hbar\omega_{p}$, respectively. The other parameters are $\Gamma_{L2}=0.01~\hbar\omega_{p}$, $\Gamma_{R1}=0.001~\hbar\omega_{p}$, $\varepsilon_{2}=0.5~\hbar\omega_{p}$, $\varepsilon_{1}=-0.5~\hbar\omega_{p}$, $m_{ep}=5\times10^{-4}~\hbar\omega_{p}$, $\gamma_{p}=1\times10^{-6}~\hbar\omega_{p}$, $\mu_{L}=0.8~\hbar\omega_{p}$, and $k_{B}T=0.1~\hbar\omega_{p}$. In our calculations, we set $e=k_{B}=\hbar=1$. }
\label{LASER}
\end{figure}

 \subsubsection{Bias dependence of the vibrational state}
 In Fig.~\ref{LASER}(a), the effective temperature $T_{\rm eff}$ and the average population $\langle n \rangle$ are plotted as a function of the chemical potential of right electrode $\mu_{R}$ with fixed  $\mu_{L}=0.8~\hbar\omega_{p}$. As we can see, the magnitudes of $T_{\rm eff}$ and $\langle n \rangle$ decrease with  increasing $\mu_{R}$ from $-1.8~\hbar\omega_{p}$ to $0.8~\hbar\omega_{p}$. The reason is as follows. By adjusting $\mu_{R}$ [the insert in Fig.~\ref{LASER}(a)], we can get two electron transport regimes and vibration statistics. For $\mu_{R}<\varepsilon_{1}$, the electron in left electrode can tunnel to level 2 and relax to level 1, accompanied by emission of a vibration. The electron in level 1 can tunnel to right electrode afterwards.
 For $\mu_{R}>\varepsilon_{1}$, the inelastic transition is blocked because the level 1 is always populated by one electron from right electrode. Due to the strong Coulomb interaction, no electron can be injected from left electrode
to level 2, such that no vibration can be excited. In such case, $T_{\rm eff}$ reduces to the  temperature of the vibration bath, see the dotted line mark in Fig.~\ref{LASER}(a).

 Similar analysis using effective temperature has been performed in previous studies\cite{chen2003local,chen2005inelastic, huang2006measurement,galperin2007heat,galperin2007molecular,huang2007local,galperin2007heat,schulze2008resonant,tsutsui2008local,schulze2008resonant,ward2011vibrational,arrachea2014vibrational,lykkebo2016single}.
 Here, we go one step further and compare the thermal $S_{\rm th}$ and the von Neumann entropy $S_{\rm vN}$ to characterize the deviation of the vibration from thermal state.  The difference of the entropy $\Delta S=S_{\rm th}-S_{\rm vN}$ indicates the nonequilibrium nature of the steady state. When they differ from each other, it is not enough to describe the vibrational state with a single effective temperature. As expected, we observe this situation in Fig.~\ref{LASER}(b). For example, when $\mu_{R}<\varepsilon_{1}$, the population inversion between level 2 and 1 leads to a vibrational lasing situation. The lasing threshold is located at $\mu_{R}=\varepsilon_{1}$. Above the threshold ($\mu_{R}<\varepsilon_{1}$), the vibration statistics obey Poisson distribution and $S_{\rm th} \neq S_{\rm vN}$ [Fig.~\ref{LASER}(c)]. Below the threshold ($\mu_{R}>\varepsilon_{1}$), the vibration reaches the thermal state, where $p_m$ follows Boltzmann distribution [Fig.~\ref{LASER}(d)] and $S_{\rm th}=S_{\rm vN}$. Therefore, a single effective temperature is only suitable for describing thermal vibrations below the threshold.

 \subsubsection{Vibration thermalization}
\begin{figure}
\includegraphics[scale=0.2]{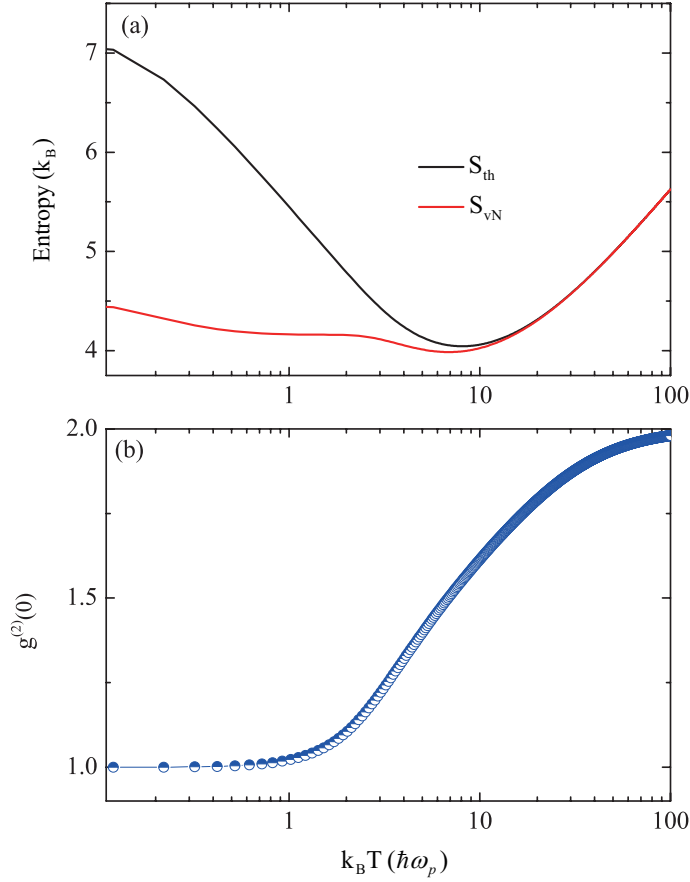}
\caption{(Color online) The effect of temperature on the entropy (a) and the second-order coherence function $g^{(2)}(0)$ (b) at $\mu_{R}=-0.8~\hbar\omega_{p}$. The other parameters are the same as in Fig.~\ref{LASER}.}
\label{tem-g2}
\end{figure}

We now consider the effect of temperature on vibration statistics, see Fig.~\ref{tem-g2}. Above, we have analyzed the range of $k_{B}T\ll\hbar\omega_{p}$. The difference between the thermal entropy and the von Neumann entropy indicates that the effective temperature is not applicable at $\mu_{R}<\varepsilon_{1}$ (above the threshold of laser). While for $k_{B}T\gg\hbar\omega_{p}$ and $\mu_{R}<\varepsilon_{1}$, one may expect $S_{\rm th}=S_{\rm vN}$. This is a consequence of thermalization of the vibrational mode due to the coupling with high temperature vibration-bath. 
To show the crossover of the vibration statistics from low temperature to high temperature more explicitly,
in Fig.~\ref{tem-g2}, we present the temperature dependence of the second-order coherence function $g^{(2)}(0)$. This clearly shows that the  transitions of the vibration state from coherent to thermal, corresponding to $g^{(2)}(0)=1$ to $g^{(2)}(0)=2$. This again shows that the effective temperature is suitable for describing thermal vibrations, but not for coherent vibrations.

Figures~\ref{LASER}-\ref{tem-g2} are the first main result of this work, showing the vibration coupled to electron weakly can reach thermal or coherent state, and that one effective temperature is not enough to describe such state. A different
way to demonstrate the effective temperature is to consider the strong electron-vibration coupling which can excite non-thermal vibrations other than the coherent states. We will discuss the nature and origin of such non-thermal vibrations in the next section.

\subsection{Results for model II}
\begin{figure}
\includegraphics[width=0.95\textwidth]{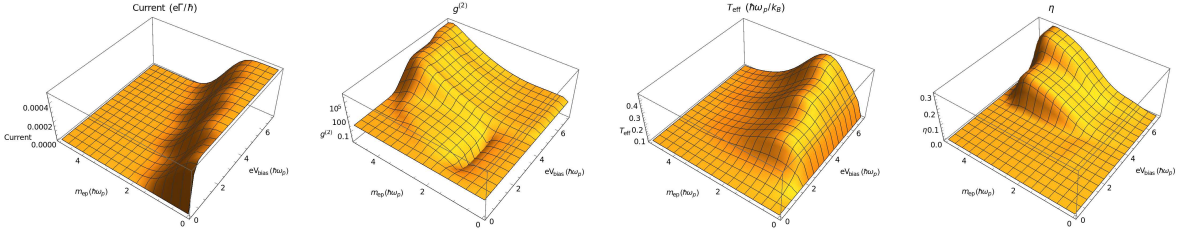}
	\caption{(Color online) 3D plot of the effective temperature $T_{\rm eff}$, the relative difference between  $S_{\rm th}$ and $S_{\rm vN}$ ($\eta$), and the second-order coherence function $g^{(2)}(0)$ as functions of the voltage bias $V_{\rm bais}$ and electron-vibration coupling constant $m_{ep}$. A symmetric voltage drop is applied to the two electrodes.
	Other parameters are $\varepsilon'=0$, $\Gamma_L =\Gamma_R =\Gamma =1\times 10^{-3}~\hbar\omega_{p}$
		, $k_{B}T=0.06~\hbar\omega_{p}$, $U'=\infty$, and $\gamma_{p}=0.01~\hbar\omega_{p}$. }
	\label{strong-single-3d}
\end{figure}

\begin{figure}
\includegraphics[width=0.99\textwidth]{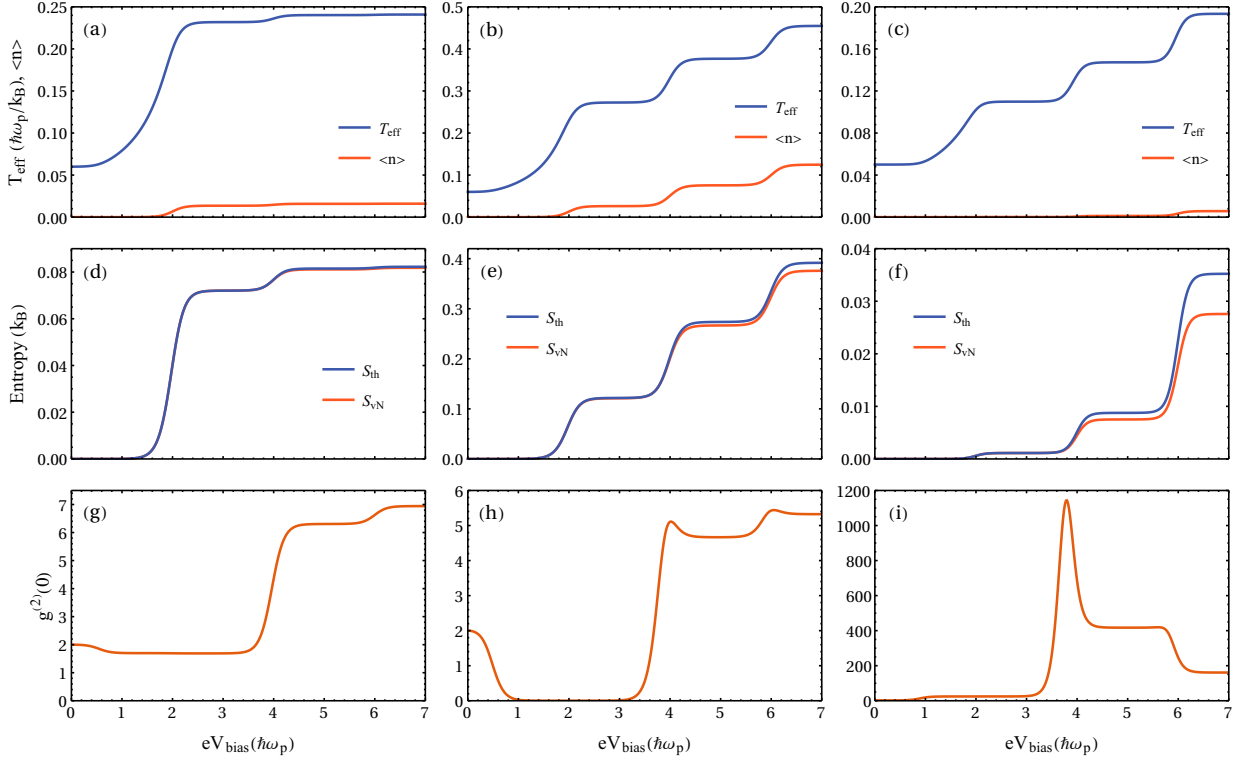}
\caption{(Color online) The effective temperature $T_{\rm eff}$, the average vibration occupation $\langle n \rangle$, the von Neumann entropy $S_{\rm vN}$, the thermal entropy $S_{\rm th}$, and the second-order coherence function $g^{(2)}(0)$ versus the voltage bias $V_{\rm bias}$, where $m_{ep}= 0.4~\hbar\omega_{p}$, $m_{ep}=1.4~\hbar\omega_{p}$ and $m_{ep}=3~\hbar\omega_{p}$ is calculated in the first, second and third row. The other parameters are the same as in Fig.~\ref{strong-single-3d}.}
\label{strong-single}
\end{figure}

Now we consider the single-level model in Fig.~\ref{model}(b). The rate equation is applied under the polaron representation by using Lang-Firsov transformation, as discussed in section \ref{Model-II}.
Figure~\ref{strong-single-3d} summarizes the dependence of $T_{\rm eff}$, the relative difference between $S_{\rm th}$ and $S_{\rm vN}$ defined as $\eta=(S_{\rm th}-S_{\rm vN})/S_{\rm th}$ and $g^{(2)}(0)$ on the voltage bias $V_{bias}$ and $m_{ep}$. Figure~\ref{strong-single} shows the line plots of their values for representative values of $m_{ep}$ for weak, medium and strong couplings.  

At low electron-vibration coupling ($m_{ep}=0.4~\hbar\omega_{p}$), the single vibration emission process is dominant, see  $T_{\rm eff}$ and $\langle n \rangle$ in Fig.~\ref{strong-single}(a). There is no obvious difference between $S_{\rm vN}$ and $S_{\rm th}$, especially in the low bias region [Fig.~\ref{strong-single}(c)]. Consequently, the effective temperature works very well. 
When the electron-vibration is increased ($m_{ep}=1.4~\hbar\omega_{p}$), multi-vibration excitation becomes possible, and Franck-Condon steps appear [Fig.~\ref{strong-single}(b)]. Anti-bunching among emitted vibrations ($g^{(2)}(0)<1$) can be observed  near the first Franck-Condon step ($eV_{\rm bias}=\hbar\omega_{p}$), which has been discussed in details in Ref.~\cite{schaeverbeke2019single}. In this regime, single vibration emission dominates. Thus, $S_{\rm vN}$ and $S_{\rm th}$ still coincide with each other. The difference between $S_{\rm vN}$ and $S_{\rm th}$ becomes obvious at larger bias [Fig.~\ref{strong-single}(e)].  Further increasing $m_{ep}$ leads to larger deviation between the two entropies at high bias [Fig.~\ref{strong-single}(f)]. 
Comparing different cases, we find that the deviation from thermal state characterized by $\Delta S= S_{\rm th} - S_{vN}$ happens at large $V_{\rm bias}$ and high $m_{ep}$, when the multi-vibration excitation process becomes important. In this case, the vibrations show super-bunching with huge $g^{2}(0)$. 
\begin{figure}[h]
\includegraphics[scale=0.6]{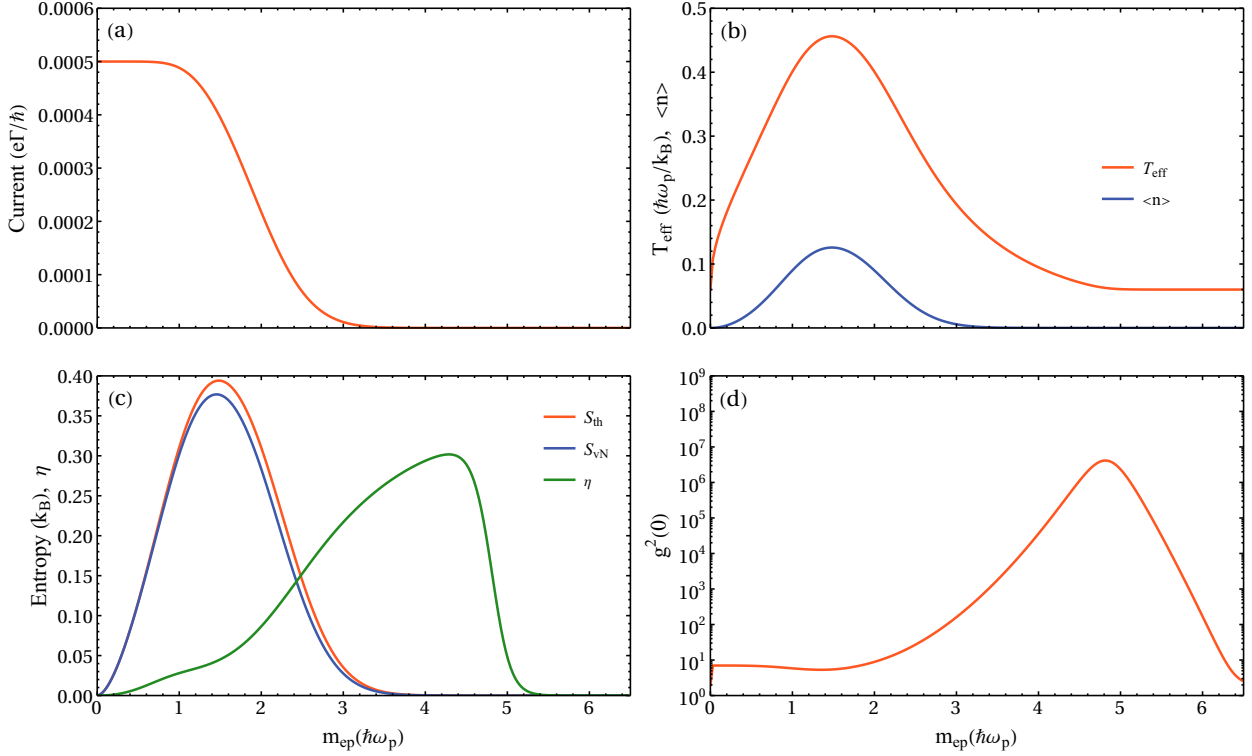}
\caption{(Color online) The current, the effective temperature $T_{\rm eff}$, the average vibration occupation $\langle n \rangle$, the von Neumann entropy $S_{\rm vN}$, the thermal entropy $S_{\rm th}$, the relative entropy difference $\eta$, and the second-order coherence function $g^{(2)}(0)$ as a function of the electron-vibration coupling strength $m_{ep}$ at $eV_{\rm bias}=7~\hbar\omega_{p}$. The other parameters are the same as in Fig.~\ref{strong-single-3d}.}
\label{rate-mep}
\end{figure}
\begin{figure}[h]
\includegraphics[scale=0.4]{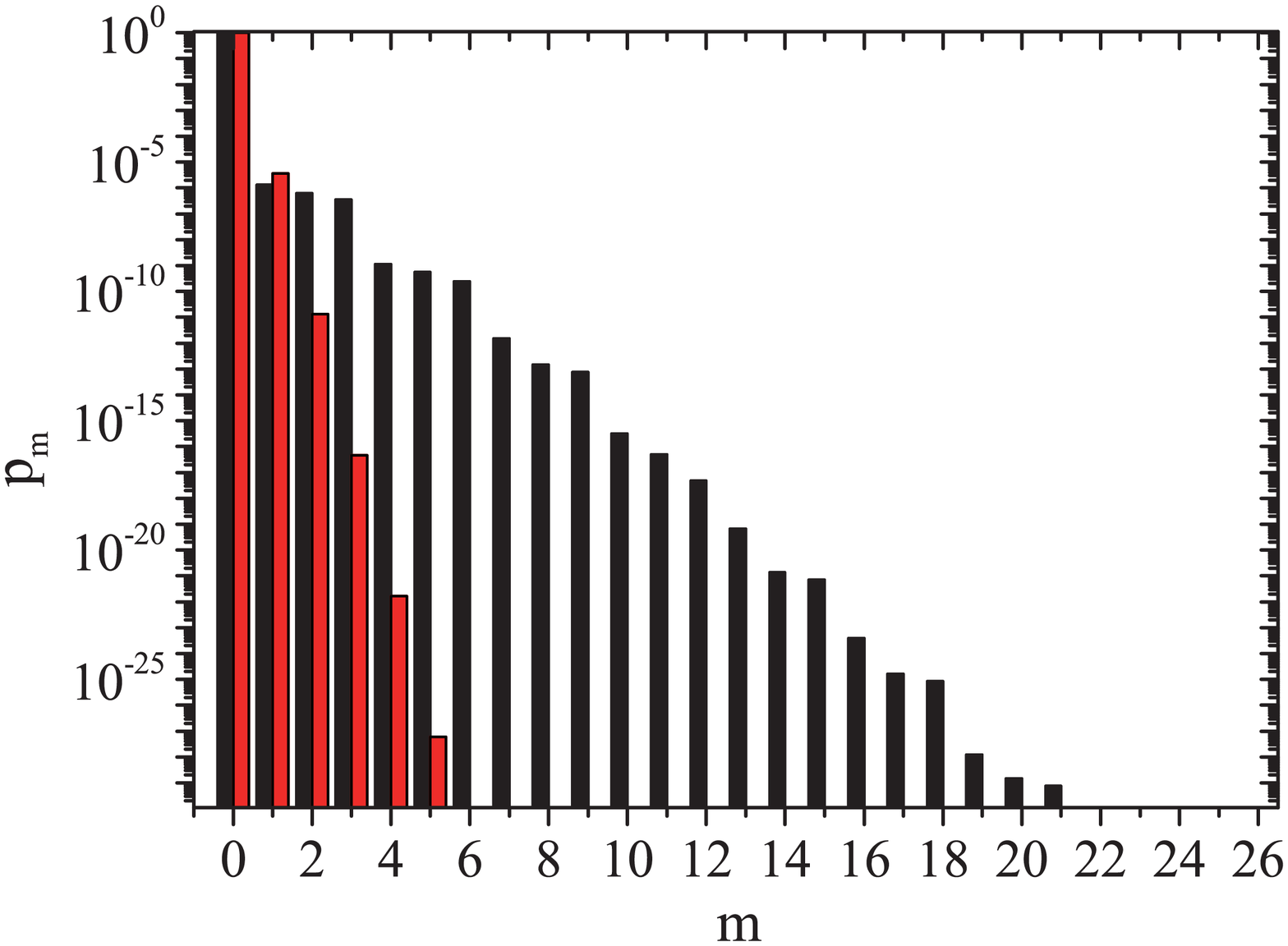}
\caption{Vibration statistics when $\eta$ reach a maximum as in Fig~\ref{rate-mep}(c). The red histogram is obtained from equilibrium distribution $p_m=e^{-m\hbar\omega_{p}/k_{B}T} (1 - e^{-\hbar\omega_{p}/k_{B}T})$ with $T=T_{\rm eff}$.}
\label{fig:pm-max}
\end{figure}

On the other hand, as shown in Fig.~\ref{strong-single-3d}, the change of $g^{(2)}(0)$, $T_{\rm eff}$ and $\eta$ with $m_{ep}$ is not monotonic. To further investigate this effect, we show $m_{ep}$ dependence of these quantities at a given bias $eV_{\rm bias}=7~\hbar\omega_{p}$ in Fig.~\ref{rate-mep}. This corresponds to line cuts of the 3D plot. In Fig.~\ref{rate-mep}(a), we can find that the current is significantly suppressed as $m_{ep}$ increases. This can be attribute to the Franck-Condon blockade, which has been discussed in Ref.~\cite{koch2005franck}. As shown in Fig.~\ref{fig:fr} of Appendix \ref{appd:frME}, when $m_{ep}=0.4~\hbar\omega_{p}$, maximum of the Franck-Condon matrix elements is near the diagonal part where the difference in vibrational occupation number between initial and final states is small. As $m_{ep}$ increases, the maximum moves away from the diagonal. Higher occupation number difference needs higher excitation energy and consequently larger voltage bias. For fixed voltage bias, increasing $m_{ep}$ results in current suppression. For $T_{\rm eff}$ or $\langle n \rangle$ in Fig.~\ref{rate-mep}(b), there exists a maximum at intermediate value of $m_{ep}\sim 1.4~\hbar\omega_{p}$. The reason is following. For one limit $m_{ep}=0$, there is no vibration excitation, such that $T_{\rm eff}=T$ and $\langle n \rangle\approx0$. For the other limit with large $m_{ep}$ Franck-Condon blockade leads to suppression of vibration excitation, again resulting in $T_{\rm eff}=T$ and $\langle n \rangle\approx0$. Thus, there exists a maximum between the two limits. Similar behavior is found for the entropy [Fig.~\ref{rate-mep}(c)]: $S_{\rm vN}=S_{\rm th}\approx 0$ for $m_{ep}\ll\hbar\omega_{p}$ and $m_{ep}\gg\hbar\omega_{p}$, corresponding thermal vibrations (see also $g^{(2)}(0)$ in Fig.~\ref{rate-mep}(d)).  The basic features of $\eta$  are similar to those of $S_{\rm th}$ and $S_{\rm vN}$. The maximum of $\eta$ moves to larger $m_{ep}$ compared to $T_{\rm eff}$ or $\langle n \rangle$. We have shown the statistical distribution of different vibrational states in Fig.~\ref{fig:pm-max}, where deviation from Boltzmann distribution can be clearly seen.

\begin{figure}
\includegraphics[scale=0.38]{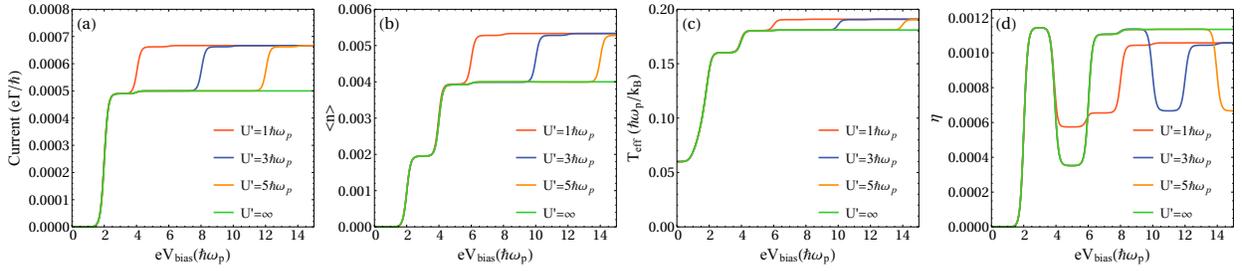}
\caption{(Color online) The current, the effective  temperature $T_{\rm eff}$, the average vibration occupation $\langle n \rangle$, and the ratio of entropy difference $\eta$ as a function of the bias $V_{\rm bias}$ for indicated values of the Coulomb interaction $U'$ at $m_{ep}=0.2~\hbar\omega_{p}$. The other parameters are the same as in Fig.~\ref{strong-single}.
}
\label{rate-u}
\end{figure}

Up to this point, we considered the strong Coulomb interaction with $U'=\infty$, where no more than one electron can reside on the molecule. For $U'<eV_{\rm bias}$ one may expect more than one electron participate the transport at the same time. Therefore, we show the effect of the Coulomb interaction on the vibration statistics in Fig.~\ref{rate-u}. We find additional Coulomb blockade steps in the results. Although $\eta$ changes at Coulomb blockade steps, the overall change is quite small and does not change much with $U'$ in the weak electron-vibration coupling regime.
\subsection{Discussions}
Energy dissipation in molecular junction has received considerable attention in the past years. It is normally termed Joule heating, although subsequent studies show that, in addition to stochastic Joule heating, electrical current can also do deterministic work on the nuclei. 
The effective temperature has been widely used to quantifying the nonequilibrium steady state of vibrations. Our results in this work show that, a single effective temperature can not always fully describe this vibrational steady state. We show that, the difference between the effective thermal entropy $S_{\rm th}$ and the actual von Neumann entropy $S_{\rm vN}$ can be used to quantify the deviation from thermal state. 
	
Since the thermal entropy is always larger than the actual entropy, their difference $\Delta S$ can be used to characterize the nonequilibrium nature of the vibrational steady state. More importantly, $\Delta S>0$ implies that the nonequilibrium free energy $F=U-TS_{\rm vN}>F_{\rm th}$. Thus, this extra free energy $\Delta F=F-F_{\rm th}$ can be used, at least in principle, freely in carefully designed thermodynamic processes. If one were to use only the effective temperature to characterized the vibrational state, one could get results that violate the second Law of thermodynamics, i.e., a Carnot engine with efficiency larger than the Carnot efficiency. This, of course, does not violate the second law, since the vibration is not in thermal equilibrium. The nonthermal statistical distribution is an extra resource that can be used to perform work.

\section{Conclusions}
In conclusion, we have presented an analysis of the vibration statistics in biased molecular junctions. By using the Lindblad master equation, a two-level molecular system with weak electron-vibration coupling was studied. It is found that the vibrational steady state before and after the lasing threshold bias are different in nature. The former can be well described by a single effective temperature, while in the latter case, the vibration is in coherent state, and an additional quantity $\Delta S$ is needed to quantify its nonequilibrium property. We also considered a single level coupling to one vibrational mode in the Holstein form. The rate equation with Lang-Firsov transformation is used to study the vibration statistics. 
The results indicate that for large electron-vibration coupling and high bias,  similar to the lasing situation in the two-level model, multi-vibration emission leads to nonequilibrium state with lower entropy and thus higher free energy. These results show that the vibration mode in biased molecular junctions can not always be characterized by a single effective temperature. The nonequilibrium vibrations may be utilized in carefully designed thermodynamic machines to achieve higher efficiencies. We considered molecular junctions in this work, but our model can be easily extended and applied to artificial molecules, i.e.,  quantum dot systems.

\section*{Acknowledgments}
This work is supported by the National Natural Science Foundation of China (Grant No. 21873033), the National Key Research and Development Program of China (Grant No. 2017YFA0403501) and the program for HUST academic frontier youth team.

\appendix
\section{Matrix elements of the density operator}
\label{appd:MEdo}
The matrix elements of the electron-vibration density operator $\rho$ can be defined as 
\begin{equation}
\rho_{m,n}^{ij}(t):=\langle m,i|\rho|j,n \rangle,
\end{equation}
where $i,j=0,g,e$ and $m/n$ is the vibration Fock state. Then, we can get the matrix elements
\begin{equation}
\begin{split}
\dot{\rho}_{m,n}^{00}&=-i\omega_{p}(m-n)\rho_{m,n}^{00}-(\Gamma_{L1}^{f}+\Gamma_{R1}^{f}+\Gamma_{L2}^{f}+\Gamma_{R2}^{f})\rho_{m,n}^{00}
+(\Gamma_{L1}^{fo}+\Gamma_{R1}^{fo})\rho_{m,n}^{gg}+(\Gamma_{L2}^{fo}+\Gamma_{R2}^{fo})\rho_{m,n}^{ee}\\
&+\frac{\gamma_{p}}{2}n_{B}[2\sqrt{m}\sqrt{n}\rho_{m-1,n-1}^{00}-(m+n+2)\rho_{mn}^{00}]\\
&+\frac{\gamma_{p}}{2}(n_{B}+1)[2\sqrt{m+1}\sqrt{n+1}\rho_{m+1,n+1}^{00}-(m+n)\rho_{mn}^{00}],
\end{split}
\end{equation}

\begin{equation}
\begin{split}
\dot{\rho}_{m,n}^{gg}&=-i\omega_{p}(m-n)\rho_{m,n}^{gg}+(\Gamma_{L1}^{f}+\Gamma_{R1}^{f})\rho_{m,n}^{00}-(\Gamma_{L1}^{fo}+\Gamma_{R1}^{fo})\rho_{m,n}^{gg}
-im_{ep}(\sqrt{m}\rho_{m-1,n}^{eg}-\sqrt{n}\rho_{m,n-1}^{ge})\\
&+\frac{\gamma_{p}}{2}n_{B}[2\sqrt{m}\sqrt{n}\rho_{m-1,n-1}^{gg}-(m+n+2)\rho_{mn}^{gg}]\\
&+\frac{\gamma_{p}}{2}(n_{B}+1)[2\sqrt{m+1}\sqrt{n+1}\rho_{m+1,n+1}^{gg}-(m+n)\rho_{mn}^{gg}],
\end{split}
\end{equation}

\begin{equation}
\begin{split}
\dot{\rho}_{m,n}^{ee}&=-i\omega_{p}(m-n)\rho_{m,n}^{ee}+(\Gamma_{L2}^{f}+\Gamma_{R2}^{f})\rho_{m,n}^{00}-(\Gamma_{L2}^{fo}+\Gamma_{R2}^{fo})\rho_{m,n}^{ee}
-im_{ep}(\sqrt{m+1}\rho_{m+1,n}^{ge}-\sqrt{n+1}\rho_{m,n+1}^{eg})\\
&+\frac{\gamma_{p}}{2}n_{B}[2\sqrt{m}\sqrt{n}\rho_{m-1,n-1}^{ee}-(m+n+2)\rho_{mn}^{ee}]\\
&+\frac{\gamma_{p}}{2}(n_{B}+1)[2\sqrt{m+1}\sqrt{n+1}\rho_{m+1,n+1}^{ee}-(m+n)\rho_{mn}^{ee}],
\end{split}
\end{equation}

\begin{equation}
\begin{split}
\dot{\rho}_{m,n}^{ge}&=-i\omega_{p}(m-n)\rho_{m,n}^{ge}-im_{ep}(\sqrt{m}\rho_{m-1,n}^{ee}-\sqrt{n+1}\rho_{m,n+1}^{gg})+i(\varepsilon_{l}-\varepsilon_{h})\rho_{m,n}^{ge}\\
&-(\frac{1}{2}\Gamma_{L1}^{fo}+\frac{1}{2}\Gamma_{R1}^{fo}+\frac{1}{2}\Gamma_{L2}^{fo}+\frac{1}{2}\Gamma_{R2}^{fo})\rho_{m,n}^{ge}\\
&+\frac{\gamma_{p}}{2}n_{B}[2\sqrt{m}\sqrt{n}\rho_{m-1,n-1}^{ge}-(m+n+2)\rho_{mn}^{ge}]\\
&+\frac{\gamma_{p}}{2}(n_{B}+1)[2\sqrt{m+1}\sqrt{n+1}\rho_{m+1,n+1}^{ge}-(m+n)\rho_{mn}^{ge}],
\end{split}
\end{equation}

\begin{equation}
\begin{split}
\dot{\rho}_{m,n}^{eg}&=-i\omega_{p}(m-n)\rho_{m,n}^{eg}-im_{ep}(\sqrt{m+1}\rho_{m+1,n}^{gg}-\sqrt{n}\rho_{m,n-1}^{ee})-i(\varepsilon_{l}-\varepsilon_{h})\rho_{m,n}^{eg}\\
&-(\frac{1}{2}\Gamma_{L1}^{fo}+\frac{1}{2}\Gamma_{R1}^{fo}+\frac{1}{2}\Gamma_{L2}^{fo}+\frac{1}{2}\Gamma_{R2}^{fo})\rho_{m,n}^{eg}\\
&+\frac{\gamma_{p}}{2}n_{B}[2\sqrt{m}\sqrt{n}\rho_{m-1,n-1}^{eg}-(m+n+2)\rho_{mn}^{eg}]\\
&+\frac{\gamma_{p}}{2}(n_{B}+1)[2\sqrt{m+1}\sqrt{n+1}\rho_{m+1,n+1}^{eg}-(m+n)\rho_{mn}^{eg}],
\end{split}
\end{equation}
where
\begin{equation}
\begin{split}
&\Gamma_{Li}^{f}=\Gamma_{Li}f_{L}(\varepsilon_{i}),\\
&\Gamma_{Ri}^{f}=\Gamma_{Ri}f_{R}(\varepsilon_{i}),\\
&\Gamma_{Li}^{fo}=\Gamma_{Li}[1-f_{L}(\varepsilon_{i})],\\
&\Gamma_{Ri}^{fo}=\Gamma_{Ri}[1-f_{L}(\varepsilon_{i})],i=1,2.
\end{split}
\end{equation}
Here, $f_{\alpha}(\varepsilon_{i})=1/[e^{(\varepsilon_{i}-\mu_{\alpha})/k_{B}T}+1]$ is the Fermi-Dirac distribution of electrode $\alpha$ with the chemical potential $\mu_{\alpha}$ and the temperature $T$.
Note that, we limit our study to the vibration laser driven by the bias voltage, such that we take $\gamma_{p}$ and $m_{ep}$ are much smaller than the molecule-electrode coupling $\Gamma_{\alpha i}$\cite{lambert2015bistable,agarwalla2019photon}.

\section{Franck-Condon matrix elements}
\label{appd:frME}
The wave function of vibration state $\ket{n}$ is given by the $n$th harmonic oscillator wave function
\begin{equation}
    \phi_{n}(x)=\left(\pi^{1 / 2} 2^{n} n! l_{\mathrm{osc}}\right){}^{-1 / 2} e^{-x^{2} /\left(2 l_{\mathrm{osc}}^{2}\right)} \mathrm{H}_{n}\left(x / l_{\mathrm{osc}}\right),
\end{equation}
in which $ l_{\rm osc} = \sqrt{\frac{\hbar}{m\omega_{p}}} $ is the oscillator length and $ \mathrm{H}_{n} $ is the hermitian polynomials.
\begin{figure}[h]
\includegraphics[scale=0.50]{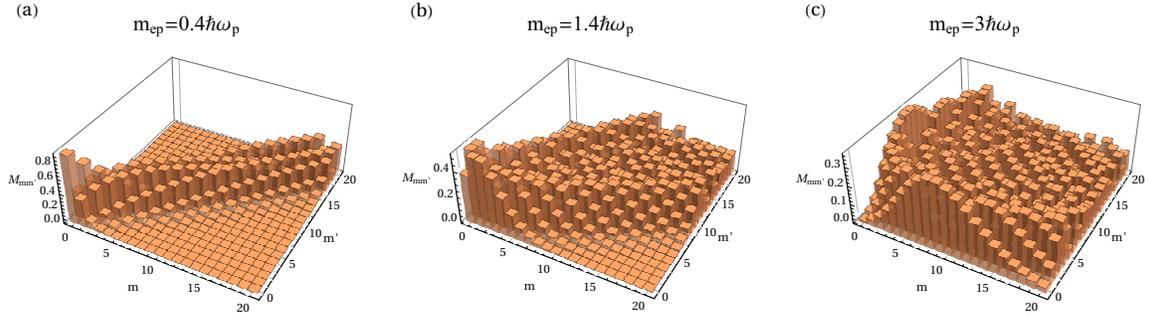}
\caption{(Color online) The Franck-Condon elements $M_{mm^{\prime}}$ for (a) $ m_{ep} = 0.4~\hbar\omega_{p}$, (b) $m_{ep} =1.4~\hbar\omega_{p}$, and (c) $m_{ep} =3~\hbar\omega_{p}$.  }
\label{fig:fr}
\end{figure}
Realizing the fact that $ e^{-\lambda(b^{\dagger}-b)} = e^{i \sqrt{2}\lambda l_{\rm osc} d/dx}$ which is the translation operator and applying the Fermi Golden rule, the Franck-Condon matrix elements can be calculated as
\begin{equation}
\begin{split}
M_{m_{1}m_{2}} &= \bra{\phi_{m_{2}}} e^{-\lambda(b^\dagger-b)} \ket{\phi_{m_{1}}}
\\
& = \braket{\phi_{m_{2}}(x) }{\phi_{m_{1}}(x - \sqrt{2}\lambda l_{osc})}
\\
& =  \left[\operatorname{sgn}\left(m_{2} - m_{1}\right)\right]{}^{m_{1} - m_{2}} \lambda^{M-m} e^{-\lambda^2 / 2}\left(\frac{m!}{M!}\right){}^{1/2} \mathrm{L}_{m}^{M-m}\left(\lambda^2\right),
\end{split}
\end{equation}
in which $\mathrm{sgn}(x)$ is the sign function, $ m = \mathrm{Min}(m_{1},m_{2})$, $ M = \mathrm{Max}(m_{1},m_{2})$ and $\mathrm{L}_{m}^{M-m}\left(\lambda^2\right)$ is the generalized Laguerre polynomials.
To show that the current suppression in Fig.~\ref{rate-mep}(a) is caused by the Franck-Condon blockade, in Fig.~\ref{fig:fr}, we plot $M_{mm^{\prime}}$ for vibration transitions from $m$ to $m^{\prime}$ with three different values of $m_{ep}$.

\bibliography{pe-one}
\end{document}